\documentclass[twocolumn,amsmath,amssymb]{revtex4}

\usepackage{graphicx}% Include figure files
\usepackage{dcolumn}% Align table columns on decimal point
\usepackage{bm}% bold math

\begin{document}
\title{Complex Probabilities on $R^N$ as Real Probabilities on $C^N$
and an Application to Path Integrals}
\author{Don Weingarten}
\altaffiliation[Present address:]{Clinton Group, 55 Water Street, 
New York, NY 10041}
\affiliation{ IBM Research, P.O. Box 218, Yorktown Heights, NY 10598}

%\receipt{27 June 2002}
\begin{abstract}

We establish a necessary and sufficient condition for averages over
complex valued weight functions on $R^N$ to be represented as
statistical averages over real, non-negative probability weights on
$C^N$. Using this result, we show that many path-integrals for
time-ordered expectation values of bosonic degrees of freedom in
real-valued time can be expressed as statistical averages over
ensembles of paths with complex-valued coordinates, and then speculate
on possible consequences of this result for the relation between
quantum and classical mechanics.

\end{abstract}

\maketitle

For a wide range of quantum mechanical systems, if real-valued time is
analytically continued to imaginary values, the weight assigned to
each path in the path integral formula for time-ordered vacuum
expectation values becomes real and non-negative, permitting
interpretation as a probability density.  Paths distributed according
to this probability density can then be generated by the Metropolis
method, one version of which makes use of the Langevin equation.

For the Langevin equation applied to paths for imaginary time
path-integrals, however, the coefficients entering the equation are
themselves analytic functions of the action occurring in the path
weight. As a consequence of this analyticity, the Langevin equation
can be analytically continued, at least formally, to complex-valued
weights \cite{Klauder}, among which are those of the original
path-integral with real-valued time.  For complex path weights, the
coefficients in the analytically continued Langevin equation and the
coordinates in the path ensembles the equation generates are complex.

Numerical tests and some analytic results concerning the Langevin
equation with complex coefficients yield a complicated picture
\cite{subsequent}. For some systems the method gives accurate answers,
for others values of the coordinates blow up as the equation's running
time is increased, and in still other applications averages over
complex Langevin trajectories converge but to wrong answers.  A
question which follows from these results is whether the failure of
the Langevin equation for many complex-valued path weights is simply a
limitation of the Langevin equation or, alternatively, arises from the
nonexistence of equivalent real, non-negative weights on
complex-valued trajectories. More generally, with no reference to the
Langevin equation, under what circumstances can averages over complex
weights on real coordinates be realized as statistical averages over
real, non-negative weights on complex coordinates?  The main result
addressing this problem so far is a proof \cite{Salcedo} that an
average over real coordinates of a complex polynomial multiplied by a
complex Gaussian can be realized as an average over a real,
non-negative weight on complex coordinates.

In the present article, by an explicit construction unrelated to the
Langevin equation, we establish a necessary and sufficient condition
for averages on $R^N$ over complex weight functions, with no explicit
restriction on functional form, to be represented as averages over
real, non-negative probability densities on $C^N$.  For the particular
case of $N = 1$, we show this condition is fulfilled for all complex
weights normalized to a total integral of 1.

As an application of our result for general $N$, we show that
real-time path integrals for many systems with bosonic degrees of
freedom can be represented as averages over ordinary statistical
ensembles of paths through complex space. Among the systems which can
be represented in this way are scalar field theories, in any
space-time dimension, with self-interaction $\phi^p$, $p$ even, field
theories, in any dimension, with bosonic degress of freedom which
range only over a bounded region, and therefore lattice gauge field
theories over any compact gauge group.  We then briefly speculate on
possible consequences of this result for the relation between quantum
and classical mechanics.

The systems we consider all have time reduced to a finite, discrete
lattice of points. We do not examine the existence of limits as the
time lattice spacing is taken to zero. We also do not discuss
algorithms for actually generating averages over the probability
densities on $C^N$ which we prove exist.

For $c(x_1, \ldots, x_N)$ a complex-valued weight function on
$(x_1, \ldots, x_N) \in R^N$,
$t(z_1, \ldots, z_N)$ a real, non-negative weight on
$(z_1, \ldots, z_N) \in C^N$, and
$p(r_1, \ldots, r_N)$ a real, non-negative weight on
$(r_1, \ldots, r_N) \in [0,\infty)^N$,
define the expectation values
\begin{subequations}
\label{Ndimcpt}
\begin{multline}
< f >_c  =
Z_c^{-1} \prod_j \left(\int_{-\infty}^{\infty}d x_j \right) \times \\
f(x_1, \ldots, x_N) \, c(x_1, \ldots, x_N)\,
\label{Ndimc}
\end{multline}
\\
\begin{multline}
< f >_t  =  Z_t^{-1}
\prod_j \left ( \int_{0}^{2 \pi} d \theta_j
\int_{0}^{\infty} r_j d r_j \right ) \times \\
f(z_1, \ldots, z_N) \, t(z_1, \ldots, z_N),
\label{Ndimt}
\end{multline}
\\
\begin{multline}
< f >_p  =
Z_p^{-1} \prod_j \left( \int_{0}^{\infty}d r_j \right ) \times \\
f(r_1, \ldots, r_N) \, p(r_1, \ldots, r_N)\,
\label{Ndimp}
\end{multline}
\end{subequations}
with $z_j$ given by $r_j \, exp( i \theta_j)$ in Eq. \ref{Ndimt}. We
require $c(x)$, $t(z)$ and $p(r)$ to have non-zero total integrals, then
choose $Z_c$, $Z_t$ and $Z_p$ to give $< 1>_c$, $< 1>_t$, $< 1>_p$ all
the value $1$. 

We will show that for any $c(x_1, \ldots, x_N)$, a necessary and
sufficient condition for the existence of a $t(z_1, \ldots, z_N)$ with
the same moments
\begin{eqnarray}
\label{Ndimmoms}
< x_1^{m_1} \ldots x_N^{m_N}>_c & = & < z_1^{m_1} \ldots z_N^{m_N}>_t
\end{eqnarray}
for all integer $m_j \ge 0$, is that
there exist a $p(r_1, \ldots, r_N)$ satisfying the bound
\begin{eqnarray}
\label{Ndimbnd}
|< x_1^{m_1} \ldots x_N^{m_N}>_c| & \le & < r_1^{m_1} \ldots r_N^{m_N}>_p
\end{eqnarray}
for all integer $m_j \ge 0$.

For notational simplicity, we consider first a detailed proof
for $N = 1$, then briefly summarize the generalization to arbitrary
$N$.

To show that Eq. \ref{Ndimmoms} implies Eq. \ref{Ndimbnd} for
some $p(r)$, we take absolute values of both sides of Eq. \ref{Ndimmoms}
and obtain
\begin{subequations}
\label{1dimabs}
\begin{eqnarray}
|Z_c^{-1} \int_{-\infty}^{\infty} x^m c(x) \, d x| & \le &
Z_t^{-1} \int_{0}^{\infty} r^m p(r) \, d r,
\label{1dimabsa}
\\
p( r) & = & r \int_{0}^{2 \pi} d \theta \, t( z).
\label{1dimabsb}
\end{eqnarray}
\end{subequations}
But $Z_p$ is identical to $Z_t$. Thus Eqs. \ref{1dimabs} implies
Eq. \ref{Ndimbnd}.

We now assume Eq. \ref{Ndimbnd} and construct a $t(z)$ which satisfies
Eq. \ref{Ndimmoms}.
For $z$ given by $r \, exp( i \theta)$ as before and a real parameter
$\lambda > 1$, Eq. \ref{Ndimbnd} implies we can
define the functions
\begin{subequations}
\label{1dimqs}
\begin{eqnarray}
q_{\lambda}( \theta ) & = &
\sum_{m \ge 0} \frac{ exp( -i m \theta) < x^m >_c}
{ \lambda^m <r^m>_p} - 1.
\label{1dimq}
\\
s_{\lambda}( \theta) & = & \{1 + 2 Re[ q_{\lambda}( \theta)]\}.
\label{1dims}
\end{eqnarray}
\end{subequations}
Eqs. \ref{1dimqs} imply for $m \ge 0$
\begin{eqnarray}
\label{1dimthetamom}
\int_0^{2 \pi} \frac{d \theta}{2 \pi} \, exp( i m \theta) \,
s_{\lambda}( \theta) & = &
\frac{ < x^m >_c}{ \lambda^m <r^m>_p}.
\end{eqnarray}
With the definitions
\begin{subequations}
\label{1dimt}
\begin{eqnarray}
< f >_{\lambda t} & = & Z_{\lambda t}^{-1} \int_0^{2 \pi}
d\theta \int_0^{\infty} d r \, r f( z) t_{\lambda}( z),
\label{1dimta}
\\
t_{\lambda} ( z) & = & r^{-1} p( {\lambda}^{-1} r) s_{\lambda}( \theta),
\label{1dimtb}
\end{eqnarray}
\end{subequations}
and $Z_{\lambda t}$ chosen to make $< 1>_{\lambda t} = 1$,
Eqs. \ref{1dimthetamom} and \ref{1dimt} then yield
\begin{eqnarray}
< z^m>_{\lambda t} & = & < x^m>_c,
\end{eqnarray}
which is Eq. \ref{Ndimmoms} for $N = 1$.

By a choice of $\lambda$ we can insure that $t_{\lambda}(z)$ is
non-negative. From Eq. \ref{1dimtb} it follows that $t_{\lambda}(z)$ is
non-negative if $s_{\lambda}(\theta)$ is non-negative. Eqs. \ref{Ndimbnd}
and \ref{1dimqs} then yield
\begin{subequations}
\begin{eqnarray}
\label{1dimsbnd}
s_{\lambda}( \theta) & \ge & 1 - 2 | q_{\lambda}( \theta)|,
\label{1dimsbnda}
\\
& \ge & 1 - 2 \sum_{m \ge 1} \frac{| < x^n >_c |}{ \lambda^m <r^n>_p},
\label{1dimsbndb}
\\
& \ge & \frac{\lambda - 3 }{\lambda - 1}.
\label{1dimsbndc}
\end{eqnarray}
\end{subequations}
Eq. \ref{1dimsbndc} imples that for $\lambda \ge 3$, $t_{\lambda}( z)$ is
non-negative.

Consider now arbitrary $N$. The proof that
Eq. \ref{Ndimmoms} implies Eq. \ref{Ndimbnd} for
some $p(r_1, \ldots, r_N)$ for arbitrary $N$ is an immediate extension of 
the
$N = 1$ proof. The proof that  Eq. \ref{Ndimbnd} implies Eq. \ref{Ndimmoms}
requires very little more effort. For general $N$, $q_{\lambda}( \theta)$
of Eq. \ref{1dimq} becomes
\begin{multline}
\label{Ndimq}
q_{\lambda}(\theta_1, \ldots, \theta_N)  =  \\
\sum_{m_i \ge 0}
\frac{ exp( -i \sum_i m_i \theta_i) < x_1^{m_1} \ldots x_N^{m_N}>_c}
{ \lambda^{\sum_i m_i} < r_1^{m_1} \ldots r_N^{m_N} >_p} - 1.
\end{multline}
Eq. \ref{1dimsbndc}, insuring the positivity of $s_{\lambda}(z_1, \ldots, 
z_N)$
and therefore of  $t_{\lambda}(z_1, \ldots, z_N)$ becomes
\begin{eqnarray}
\label{Ndimsbnd}
s_{\lambda}( \theta_1, \ldots, \theta_N) & \ge &
3 - \frac{2 \lambda^N}{(\lambda - 1)^N},
\end{eqnarray}
the right side of which is positive for
\begin{eqnarray}
\label{Ndimlambda}
\lambda & \ge & \frac{1}{1 - (\frac{2}{3})^{\frac{1}{N}}}.
\end{eqnarray}

Returning to the case of $N = 1$, we now show that a
$p(r)$ exists fulfilling Eq. \ref{Ndimbnd} for any $c(x)$ with
$<1>_c$ of $1$\cite{proof}. To do this, we make use of a theorem
giving conditions for the existence
of a $p(r)$ such that for all non-negative integers $m$
\begin{eqnarray}
< r^m >_p = s_m,
\label{momprob}
\end{eqnarray}
for a specified sequence of real numbers $s_0, s_1, \ldots$.
From the $s_m$, for any non-negative integer $n$, define the 
$(n + 1) \times (n + 1)$ matrices 
\begin{subequations}
\begin{eqnarray}
\label{defH}
H^{2 n}_{i j} & = & s_{i + j},
\label{defHa}
\\
H^{2 n + 1}_{i j} & = & s_{i + j + 1},
\label{defHb}
\end{eqnarray}
\end{subequations}
with $0 \le i, j \le n$. 
Then according to a result of Stieltjes \cite{Akhiezer}
a $p(r)$ obeying Eq. \ref{momprob} exists if for all non-negative $n$
\begin{eqnarray}
\label{posH}
det(H^n) > 0.
\end{eqnarray}
For any $c(x)$, we will construct a sequence $s_0, s_1, \ldots,$
obeying both Eqs. \ref{Ndimbnd} and \ref{posH}. Choose $s_0$ to be $1$.
Now suppose $s_0, s_1, \ldots, s_k,$ have been found obeying
Eqs. \ref{Ndimbnd} and \ref{posH} for all $n \le k$. By isolating
the dependence of $det(H^{k+1})$ on $s^{k+1}$, 
it is not hard to show that a positive real bound $b$ exists, such that
Eq. \ref{posH} is fulfulled for $n = k + 1$ for any $s_{k+1} > b$.
We choose $s_{k+1}$ to be the greater of $|<x^{k+1}>_c|$ and $b$.
By induction, the resulting sequence $s_0, s_1, \ldots$ obeys
Eqs. \ref{Ndimbnd} and \ref{posH} for all $m$. It follows that a
$p(r)$ exists fulfilling Eq. \ref{momprob} and thus also 
Eq. \ref{Ndimbnd}. 

The result we have just proved can not be extended automatically to $N
> 1$ since, for $N > 1$, the analog of Eq. \ref{posH} is not a sufficient
condition for the existence of a $p(r_1, \ldots, r_N)$ fulfilling the
analog of Eq. \ref{momprob} \cite{Berg}.

For each $c( x_1, \ldots, x_N)$ obeying Eq. \ref{Ndimbnd} for some $p(
r_1, \ldots, r_N)$, the non-negative, real $t_{\lambda}(z_1, \ldots,
z_N)$ we construct satisfying Eq. \ref{Ndimmoms} is only one
representative of a large family of such functions fulfilling
Eq. \ref{Ndimmoms}. At the very least, given any $p( r_1, \ldots,
r_N)$ obeying Eq. \ref{Ndimbnd}, an infinite family of distinct $p'(
r_1, \ldots, r_N)$ which also obey Eq. \ref{Ndimbnd} can be
constructed by shifting some part of the weight carried by $p( r_1,
\ldots, r_N)$ out to larger values of $r_1, \ldots, r_N$.  For
example, for positive $a_1, \ldots, a_N$, define $p'( r_1, \ldots,
r_N)$ to be zero except on the region $r_1 \ge a_1, \ldots, r_N \ge
a_N$, where it is given by $p( r_1 - a_1, \ldots, r_N - a_N)$.  Each
moment of $p'( r_1, \ldots, r_N)$ is greater than the corresponding
moment of $p( r_1, \ldots, r_N)$, and thus satisfies
Eq. \ref{Ndimbnd}. By means of Eqs. \ref{Ndimq} - \ref{Ndimlambda},
$p'( r_1, \ldots, r_N)$ can therefore be turned into a new
$t'_{\lambda}(z_1, \ldots, z_N)$ with moments satisfying
Eq. \ref{Ndimmoms}.

An expression for $q_{\lambda}(\theta_1, \ldots, \theta_N)$ somewhat
simpler than Eq. \ref{Ndimq} can be found if $p( r_1, \ldots,
r_N)$ in Eq.  \ref{Ndimbnd} is a product $\hat{p}(r_1) \ldots
\hat{p}(r_N)$ with moments $< r^m>_{\hat{p}}$ which, for any
$R > 0$ and integer $m \ge 0$, obey
\begin{eqnarray}
\label{phat}
< r^m>_{\hat{p}} & \ge & b_R \, R^m
\end{eqnarray}
for some $b_R > 0$ independent of $m$.
Then with the definition
\begin{eqnarray}
\label{deff}
f_{\lambda}( y) & = & \sum_{m \ge 0} \frac{y^m}{\lambda^m <r^m>_{\hat{p}}},
\end{eqnarray}
Eq. \ref{Ndimq} takes the form
\begin{multline}
\label{Ndimqfac}
q_{\lambda}(\theta_1, \ldots, \theta_N)  =  \\
< f_{\lambda}[ x_1 exp( - i \theta_1)] \ldots
f_{\lambda}[ x_N exp( - i \theta_N)] >_c - 1.
\end{multline}

For many quantum mechanical systems, a finite volume, lattice
approximation to the path integral in real-valued time leads to an
expectation value fulfilling Eq. \ref{Ndimbnd} and therefore
respresentable as an ordinary statistical average over an ensemble of
random paths through complex space. Among the systems which obey
Eq. \ref{Ndimbnd} for some $p(x_1, \ldots, x_N)$ are scalar field
theories, in any space-time dimension, with self-interaction $\phi^p$,
$p$ even, field theories, in any dimension, with bosonic degrees of
freedom which range only over a bounded region, and therefore lattice
gauge field theories over any compact gauge group.  In addition, of
course, for any system for which the real-valued time path integral is
correctly handled by the complex Langevin equation, the probability
distribution $t(z_1, \dots, z_N)$ given by the limit as running time
becomes large of the ensemble of points in $C^N$ struck by the
Langevin trajectory fulfills Eq. \ref{Ndimmoms}.  Since
Eq. \ref{Ndimmoms} implies Eq. \ref{Ndimbnd}, such a system then has a
$p(r_1, \ldots, r_N)$ fulfilling Eq. \ref{Ndimbnd}.

As an example of a proof of Eq. \ref{Ndimbnd},
consider the path integral for a single anharmonic
oscillator with time reduced to a periodic lattice of points
labeled by a positive integer $t \le T$. With $x_t$ the position
at time $t$, the
time-ordered expectation of a polynomial
$f( x_1, \ldots, x_T)$ becomes $< f>_c$ of Eq. \ref{Ndimc} with
\begin{multline}
\label{pathint}
c(x_1, \ldots, x_T) = \\ exp\left\{ i
\sum_t \left[\mu \frac{(x_t - x_{t+1})^2}{2 \delta}
- \kappa x_t^4 \delta \right]\right\},
\end{multline}
where $\mu$ is the particle's mass, $\kappa$ the anharmonic spring
constant, and $\delta$ the lattice spacing.  Rotating each integral
over real $x_t$ to an integral over $exp( -i \pi/8) y_t$ with $y_t$
real, taking the absolute value of both sides of the rotated
version of Eq. \ref{Ndimc}, and using the inequality
\begin{eqnarray}
\label{yineq}
(y_t - y_{t+1})^2 & \le & 2 ( y_t^2 + y_{t+1}^2),
\end{eqnarray}
we obtain
\begin{subequations}
\label{pathintbnd}
\begin{eqnarray}
|< x_1^{m_1} \ldots x_N^{m_N}>_c| & \le &
\frac{ Z_{\hat{p}}^T}{|Z_c|} \prod_t < r_t^{m_t}>_{\hat{p}},
\label{pathintbnda}
\\
\hat{p}( r) & = &
exp\left(\frac{\sqrt{2} r^2}{\mu \delta} - \kappa r^4 \delta \right),
\label{pathintbndb}
\end{eqnarray}
\end{subequations}
with $< \ldots>_{\hat{p}}$ given by Eq. \ref{Ndimp} for $N = 1$.
Eq. \ref{pathintbnda} then gives Eq. \ref{Ndimbnd} with
\begin{subequations}
\label{pathintbnd2}
\begin{eqnarray}
p( x_1, \ldots, x_T) & = & \prod_t \hat{p}(\frac{x_t}{ \lambda}),
\label{pathintbnd2a}
\\
\lambda & = & Max [ 1, \frac{Z_{\hat{p}}}{|Z_c|^{\frac{1}{T}}}].
\label{pathintbnd2b}
\end{eqnarray}
\end{subequations}

The proof of Eq. \ref{Ndimbnd} for an anharmonic oscillator extends
easily to a $\phi^p$ field theory, $p$ even, for any dimension of
space-time and to bosonic field theories, in any dimension, with with
degress of freedom bounded by some positive real $B$.  In a proof of
Eq. \ref{Ndimbnd} for bounded degrees of freedom, $\hat{p}(r)$ in
Eq. \ref{pathintbndb} becomes $\delta( r - B)$.

Finally, we speculate briefly on the possibility that the degrees of
freedom of the real world might actually be complex-valued,
distributed according to a probability which, as we have shown
possible in many cases, yields averages agreeing with those of
conventional quantum mechanics. A version of this idea suggested by
the complex Langevin equation is considered in Ref. \cite{Migdal}. A
problem encountered immediately by this proposal is that the
macroscopic world, governed nearly by classical mechanics, exhibits
only real-valued positions. This problem would be solved if, within
the statistical ensemble of paths through complex-space, the paths for
macroscopic variables happened nearly to obey classical mechanics and
to have only small imaginary parts. A formulation of quantum mechanics
incorporating this feature would, potentially, be capable of resolving
many of the puzzles in the interpretation of quantum mechanics
\cite{bell,nuovocim}. A discussion of the proposed resolution of these
problems by decoherence criteria \cite{nbi,others} and of the
difficulties which these encounter appears in Ref. \cite{goldstein}.

The speculation we offer here, in effect, is that quantum mechanics is
a version of ordinary statistical mechanics but for paths through a
complex-valued coordinate space. Although this prospect may sound
implausible, the harmonic oscillator provides a simple model of how
such behavior might conceivably come about.

Assume periodic boundary conditions for time period $T = 2 M + 1$
and Fourier transform the oscillator coordinate $x_t$ at integer time $t$
according to the convention
\begin{equation}
\label{fourier}
x_t  =  \sum_{0 \le k \le M} \left[ a_k cos( \frac{ 2 \pi k t}{T}) +
b_k sin( \frac{ 2 \pi k t}{T}) \right],
\end{equation}
with real coefficients $a_k$ and $b_k$, $b_0$ identically 0. The
time-ordered expectation of a polynomial
$f(a_0, \ldots, a_M, b_1, \ldots, b_M)$ is $<f>_c$ of Eq. \ref{Ndimc} for
\begin{multline}
c(a_0, \ldots, a_M, b_1, \ldots, b_M)  = \\ exp\left[ i
\sum_{0 \le k \le M} ( a_k^2 + b_k^2) s_k \right],
\label{harmc}
\end{multline}
where $s_k$ is
\begin{equation}
s_k  =  \frac{2 \mu}{\delta}[ 1 - cos(k)] - \kappa \delta,
\label{harmsk}
\end{equation}
and $\delta$ is the time lattice spacing.

In Eq. \ref{Ndimc} expressed as integrals over real $a_k$ and $b_k$,
for $s_k$ positive the integrals can be rotated, respectively, to
$\sqrt{i} c_k$ and $\sqrt{i} d_k$ with real $c_k$ and $d_k$. For $s_k$
negative, the $a_k$ and $b_k$ integrals can be rotated, respectively,
to $\sqrt{-i} c_k$ and $\sqrt{-i} d_k$. If $\mu$, however, is given a
microscopic positive imaginary part $\epsilon$ and a correctly chosen
real part, there will be a single $k$ for which $s_k$ has no real part
and a microscopic positive imaginary part. For this $k$ the integrals
over $a_k$ and $b_k$ can be left pure real. Inverting the Fourier
transform of Eq. \ref{fourier}, a statistical ensemble of
complex-valued trajectories $x_t$ results with real, non-negative
probability weight.  A typical trajectory in this ensemble consists of
the sum a real part, with amplitude of order $\epsilon^{-1/2}$,
oscillating at the frequency predicted by classical mechanics, and a
complex part independent of $\epsilon$.

The preceding construction is, effectively, an application to the
harmonic oscillator path integral of the method of steepest descent.
A similar construction can be carried out for a free scalar field
theory in any dimension of space time. For more complicated field
theories, the method of steepest descent can also be applied to obtain
a $t( x_1, \ldots, x_N)$ which includes real-valued trajectories
obeying the classical equations of motion, but the $t(x_1, \ldots,
x_N)$ produced in this way will only approximately fulfill
Eq. \ref{Ndimmoms}. Whether for interacting systems $t(x_1,
\ldots, x_N)$ can be found which fulfill Eq. \ref{Ndimmoms} and
include real-valued trajectories obeying classical equations of
motion is an open question.

\end{document}